\documentclass[12pt]{article}

\usepackage[parfill]{parskip}
\usepackage{setspace}
\usepackage[round]{natbib}
\usepackage{amsmath,amssymb}
\usepackage{graphicx}

\newcommand{\be}{\begin{equation}}
\newcommand{\ee}{\end{equation}}
\newcommand{\beq}{\begin{eqnarray}}
\newcommand{\eeq}{\end{eqnarray}}

\newcommand{\bN}{{\mathbf{N}}}
\newcommand{\bR}{{\mathbf{R}}}

\newcommand{\bQ}{{\mathbb{Q}}}
\newcommand{\bx}{{\mathbf{x}}}
\newcommand{\sR}{\mathcal{R}}
\newcommand{\nn}{\nonumber}
\newcommand{\bNb}{\mathbf{n}^{\mathrm{  B}}}
\newcommand{\bNt}{\mathbf{n}^{\mathrm{T}}}
\newcommand{\bRb}{\mathbf{r}^{\mathrm{B}}}
\newcommand{\bRt}{\mathbf{r}^{\mathrm{T}}}
\newcommand{\bFb}{\mathbf{F}^{\mathrm{B}}}
\newcommand{\bFt}{\mathbf{F}^{\mathrm{T}}}

\newtheorem{theorem}{Theorem}

\begin{document}

\begin{titlepage}

{ \bf \noindent Inferring Species Trees Directly from Biallelic Genetic Markers: Bypassing Gene Trees in a Full Coalescent Analysis}\\

\noindent Research article

David Bryant\\ {\em Department of Mathematics and Statistics, University of Otago, Dunedin, New Zealand, and the Allan Wilson Centre for Molecular Ecology and Evolution} \\

Remco Bouckaert\\ {\em Computational Evolution Group, University of Auckland, Auckland, New Zealand}\\

Joseph Felsenstein\\ {\em Department of Genome Sciences and Department of
Biology, University of Washington, Box 355065, Seattle, WA 98195-5065}\\

Noah A. Rosenberg\\ {\em Department of Biology, Stanford University, Stanford, California, USA.}\\

Arindam RoyChoudhury\\ {\em Mailman School of Public Health, Columbia University, New York, USA.}\\

{\bf Corresponding Author:} David Bryant, Department of Mathematics and Statistics, University of Otago, P.O. Box 	56, Dunedin, New Zealand.\\
ph: +64 3 4797889. fax +64 3 479 8427.\\
 email: {\tt david.bryant@otago.ac.nz}\\


{\bf Running head:} Inferring Species Trees from Markers
\end{titlepage}
\doublespacing

\pagebreak
\begin{abstract}
The multi-species coalescent provides an elegant theoretical framework for estimating species trees and species demographics from genetic markers. Practical applications of the multi-species coalescent model are, however, limited by the need to integrate or sample over all gene trees possible for each genetic marker. Here we describe a polynomial-time algorithm that computes the likelihood of a species tree directly from the markers under a finite-sites model of mutation, effectively integrating over all possible gene trees. The method applies to independent (unlinked) biallelic markers such as well-spaced single nucleotide polymorphisms (SNPs), and we have implemented it in SNAPP, a Markov chain Monte-Carlo sampler for inferring species trees, divergence dates, and population sizes. We report results from simulation experiments and from an analysis of 1997 amplified fragment length polymorphism (AFLP) loci in 69 individuals sampled from six species  of {\em Ourisia} (New Zealand native foxglove). 
\end{abstract}

\pagebreak

\section*{Introduction}

Biallelic markers such as single nucleotide polymorphisms (SNPs) and amplified fragment length polymorphisms (AFLPs) are potentially rich sources of information about species radiations, species divergences and historical demographics. However, extracting  this information is not always straightforward. Patterns of genetic variation at these markers are 
not just a product of the relationships between the species; they also reflect  
 inheritance patterns within each species. Any full-likelihood (or full-Bayesian) method for inferring species histories from genetic markers  needs to model the random distribution of gene tree histories for each marker. To date, this task has often meant implementing massive Monte-Carlo simulation-based sampling of not only species trees, but also the gene trees at each locus \citep{Rannala03,Wilson03,Liu07,Hey07,Heled10}. 

In this paper,  we describe an algorithm that allows us to bypass the gene trees and compute species tree likelihoods directly from the markers. The likelihood values, or posterior probabilities, computed by the algorithm are identical to those that would be obtained by sampling every possible gene tree, and every possible set of branch lengths, at each locus. The  algorithm makes use of new formulae for lineage and allele probabilities under the coalescent \citep{Bryant11cc}, and of recently developed numerical techniques to evaluate these formulae. 

Our approach makes the following  assumptions regarding the data:
\begin{enumerate} 
\item[(A1)] Each marker is a single biallelic character (e.g. a biallelic SNP or AFLP banding pattern);
\item[(A2)] The genealogies for separate markers are conditionally independent given the species tree. In practice this assumption applies to unlinked or loosely linked markers. \end{enumerate}
This  latter assumption is clearly not valid for sites in a single gene sequence. However, it is satisfied for SNPs that are well-spaced along the genome. If the independence assumption (A2) is only partially violated, the effect of linkage could be investigated by subsampling sets of markers with varying degrees of independence. 

 In principle, the full-likelihood methods of \cite{Liu07} and \cite{Heled10} could be applied to data satisfying (A1) and (A2) by encoding each marker as a separate locus. This strategy would quickly become infeasible as the number of markers increased.  \cite{Nielsen98a} demonstrated that a full-likelihood approach is tractable for data satisfying (A1) and (A2), presenting an algorithm that uses biallelic characters directly to compute the likelihood of the species tree. However, their algorithm did not permit mutation, and it was  computationally feasible only for small species trees.  \cite{RoyChoudhury06} and \cite{RoyChoudhury08} made a substantial advance on the computational problem. They took the approach of \cite{Nielsen98a} and placed it within a dynamic programming framework, thereby giving an efficient algorithm for computing the likelihood of a tree with an arbitrary number of species, but no mutation. Here we extend the dynamic programming structure of \cite{RoyChoudhury08} to incorporate mutation. We address the extensive algorithmic and mathematical challenges resulting from this deceptively minor change in model assumptions. 

We have implemented a new likelihood algorithm and incorporated it within a Bayesian Markov chain Monte-Carlo (MCMC) sampler  which we call SNAPP ({\em SNP and AFLP Phylogenies}).  SNAPP, which interfaces with the BEAST package \cite{Drummond07}, takes a range of biallelic data types as input and returns a sample of species trees with (relative) divergence times and population sizes. We have tested and validated the algorithm and the software using a range of techniques, and we report on results of two experiments with simulated data. 
The software is open source, and is available for download from
\begin{quotation}
 {\tt http://www.maths.otago.ac.nz/software/snapp}
\end{quotation}

To illustrate the application of SNAPP, we analyse AFLP loci in 69 individuals sampled from six species  of New Zealand {\em Ourisia}, or native foxglove. The New Zealand {\em Ourisia} form a relatively recent species radiation, and inference of branching patterns between these species has proven difficult, arguably due to the presence of ancestral lineage sorting \citep{Meudt09}. Our Bayesian analysis provides a relatively clear picture of ancestral species relations in the group and, up to a scale constant, effective population sizes. 

\section*{Methods}

\subsection*{The multi-species coalescent}
 
Our models are all based on the assumption that the lineage dynamics within  populations (or species) are well described by the conventional Wright-Fisher model. The distribution of the gene trees within each population is approximated by the {\em coalescent process} (reviewed in \cite{Felsenstein04,Hein05,Wakeley09}). This process models the number of ancestral lineages of the sample from a single population as a Markov process that goes {\em backwards} in time. Initially, the number of ancestral lineages equals the size of the sample. Going backwards in time (upwards in a branch) lineages meet at common ancestors, and the number of ancestral lineages decreases. 

It is customary in coalescent theory to rescale time in terms of effective population size, so that two lineages coalesce at rate $1$. This rescaling is not generally possible in the multi-species coalescent since different species can have different effective population sizes. Instead we adopt the standard practice from phylogenetics and rescale time in terms of expected mutations (as in \cite{Rannala03}). Hence the expected time to a coalescence for two lineages is $\theta/2$ and the expected time to a coalescence for $k$ lineages is $\frac{\theta}{k(k-1)}$, where $\theta$ denotes the expected divergence between any two individuals in the population.

At the first coalescent event, two lineages are selected at random and combined, and we are left with $k-1$ lineages. This coalescence of lineages continues until we reach the top of the branch, at which anywhere from 1 to $k$ lineages could be present.

 The nodes in the {\em species} tree represent species divergences, or population splits. The individuals in each of the child populations are descendants of individuals in the parent population. In terms of the coalescent process, the lineages coming upwards from the child population become lineages at the base of the parent population. This process continues upwards in the species tree until we reach the species tree root. At this point, any remaining lineages coalesce according to the standard single-population coalescent model. 

See \cite{Felsenstein04}, \cite{Degnan09}, and \cite{Heled10} for  general introductions to the multi-species  coalescent. Early contributions to the development of multi-species models built on the branches of a species tree were made by \cite{Tajima83}, \cite{Hudson83}, \cite{Takahata85}, \cite{Nei87}, \cite{Pamilo88} and \cite{Takahata89}.

The multi-species coalescent determines a distribution for gene trees and their branch lengths, conditional on a species tree. The parameters of the distribution are the shape of the species tree, the divergence times within the species tree and the population sizes along the branches of the species tree (one parameter for each branch). We  bundle these parameters into the single composite parameter $S$, so that the probability of a gene tree $G$ given the species tree is $P(G|S)$. We treat this quantity as a density rather than a discrete probability because of  the continuous branch lengths of $G$. 

Let $X$ denote the alignment of sequences for a locus. Conventional phylogenetic models (e.g. \citet{Felsenstein04}) give us the probability that $X$ evolved along a specified gene tree $G$. These models provide the distribution of states at the root and the mutation probabilities down the edges of the tree and hence determine $P(X|G)$, the probability of the data (alignment) given the gene tree. Note that once the gene tree is chosen, the species tree has no further influence on the  probability  of the data. 

Putting $P(G|S)$ and $P(X|G)$ together, we obtain the {\em joint} probability (or density) of the alignment $X$ and the gene tree $G$:
\begin{equation}
P(X,G | S) = P(X | G) P(G|S).
\end{equation}
The gene tree $G$ is  not observed directly and it can be difficult to estimate. Since our focus is on the species tree and the features of the species tree, we work with the {\em marginal} probability of the data, summing over all possible gene tree topologies and integrating over the branch lengths:
\begin{equation}
P(X|S) = \int P(X|G) P(G|S) dG. \label{eq:Felsenstein}
\end{equation}
Eq.~\eqref{eq:Felsenstein} is sometimes called the {\em Felsenstein equation} \citep{Felsenstein88,Hey07}. 
 
Generally, we consider multiple genetic markers.  We assume that the gene trees for each marker are independent.  Let $X_i$ be the alignment for the $i$th gene and let $G_i$ be a corresponding gene tree. Under the assumption that the alignments are indeed independent, the total probability of the $m$ alignments at $m$ genes is a product over all of the genes:
\begin{equation}
P(X_1,X_2,\ldots,X_m | S) = \prod_{i=1}^m P(X_i | S) = \prod_{i=1}^m \int P(X_i | G_i) P(G_i | S) dG_i. \label{eq:likelihood}
\end{equation}
If we were to plug this formula into a Bayesian analysis, we would specify a prior distribution $P(S)$ on the species trees, and then sample from the posterior distribution
\begin{equation}
P(S|X_1,\ldots,X_m) \propto \left( \prod_{i=1}^m  \int P(X_i | G_i) P(G_i | S) dG_i \right) P(S). \label{eq:posterior}
\end{equation}

Sampling from $P(S|X_1,\ldots,X_m)$ is equivalent to sampling from the joint posterior distribution
\begin{equation}
P(S, G_1,\ldots,G_m | X_1, \ldots, X_m) \propto  \left( \prod_{i=1}^m P(X_i | G_i) P(G_i | S) \right) P(S) 
\end{equation}
and only considering the marginal distribution of the species trees $S$. This is the approach taken by 
 BATWING \citep{Wilson03}, BEST \citep{Liu07} and STAR-BEAST \citep{Heled10}. Note that if the actual gene trees $G_i$ are provided, or if they can be inferred with high accuracy, they can be treated as data and the species tree can be inferred directly \citep{Degnan05,Kubatko09}.

At this point it is appropriate to reflect on what exactly is required when applying \eqref{eq:likelihood} or \eqref{eq:posterior} to large numbers of independent biallelic markers. To evaluate the likelihood exactly, we would need to sum (or integrate) over all possible gene trees. In a Bayesian setting, we would need to sample over a space containing not only every possible choice of species tree, but also every possible choice of gene tree for every locus. Furthermore, the marginal probabilities for the gene trees depend not only on the data but on the species tree, and so the analyses for the separate genes are all interdependent. An analysis of 1000 independent loci then amounts to 1001  inter-linked Bayesian analyses (1000 gene trees and one species tree). Even with modern Monte-Carlo algorithms, this type of analysis is computationally daunting. 

\subsection*{Overview of the likelihood algorithm}

We circumvent the computational difficulties by calculating the integral in \eqref{eq:likelihood} analytically. In the following sections we describe a pruning algorithm that we use to compute the  likelihood of a species tree given genotype data at unlinked biallelic markers. The algorithm works in a similar manner to Felsenstein's pruning algorithm \citep{Felsenstein81} for computing the likelihood of a gene tree:  we define partial likelihoods that focus only on a specific subtree; the partial likelihoods are then computed starting at the leaves (of the species tree), working upwards to the root.

There are two major differences. In Felsenstein's pruning algorithm, one partial likelihood is defined for every node and every state (i.e. amino acid or nucleotide). In our algorithm we have separate partial likelihoods for the top and bottom of each branch in the species tree, for every possible number of ancestral lineages at each point, and for every possible count of the number among these lineages carrying each allele. 

Secondly, we need to deal with the complication that the coalescent process works backwards in time (and is not reversible) while the mutation process works forward in time. We were not able to define a simple transition process taking numbers of ancestral lineages to numbers of descendant lineages. Instead we first compute probability distributions for the numbers of ancestral lineages at each node in the species tree. We then define partial likelihoods for subtrees in the species tree and derive the equations required to compute them efficiently. Finally, we show how to handle the probabilities at the root of the species tree when computing the full probability of the genotype data of a marker. 

We orient  trees so that the ancestral nodes are at the top and time travels downwards. Thus, the base of a branch in the species tree corresponds to the population at the time nearest to the present, while the top of a branch corresponds to the population just after it has diverged from its ancestral population. In a similar fashion, the genotypic state in a gene tree will evolve from the top of the gene tree (the common ancestor) downwards to the leaves.

\subsection*{Red and green alleles}

The multi-species coalescent model for the evolution of markers (SNPs, AFLPs etc.) has two components: the model for the gene trees in the species tree, and the model for the markers evolving down the gene tree (that is, forward in time). The model for gene trees uses a coalescent process that works backwards in time whereas the mutation model for genetic markers (SNPs, AFLPs, etc.) typically works forwards in time.

Given a gene tree with branch lengths specified, we model the evolution of a genetic marker using  standard phylogenetic machinery. Suppose that there are two alleles, which we label {\em red} and {\em green}. Let $u$ be the rate of mutation from the red allele to the green allele per unit time  (forward in time), and let $v$ be the corresponding rate of mutating from green to red. We say that a lineage is a red lineage if it has the red allele and a green lineage otherwise.

The allele of the most recent common ancestor (MRCA) at the root of the gene tree is red with probability $\pi = \frac{v}{u+v}$ and green with probability $(1-\pi) = \frac{u}{u+v}$. The marker evolves down the gene tree as a continuous-time Markov chain whose instantaneous rate matrix has rate $u$ of mutating from red to green and rate $v$ for mutating from green to red. The alleles at the leaves of the gene tree are then the observed alleles. The probability of the allele frequencies at a marker, given the species tree, is therefore the probability of the site given a gene tree multiplied by the probability of the gene tree given the species tree, summed over all possible gene tree topologies and integrated over all possible gene tree branch lengths (eq. \eqref{eq:likelihood}).

\subsection*{Ancestral lineage counts and the likelihood}

The multi-species coalescent can be used to generate a random gene tree conditional on a species tree. If we take any node or point in the species tree, we can count the number of lineages in the gene tree in that species at that point in time. We say that at a specified time point, this quantity is the number of {\em ancestral lineages}. The count of ancestral lineages is a random variable with distribution determined by the multi-species coalescent process and its resulting distribution of gene trees. The first step in our likelihood algorithm is the calculation of these lineage count distributions. See \cite{RoyChoudhury08} and \cite{Efromovich08} for similar computations.

\begin{figure}[htbp] 
   \centering
   \includegraphics[width=5in]{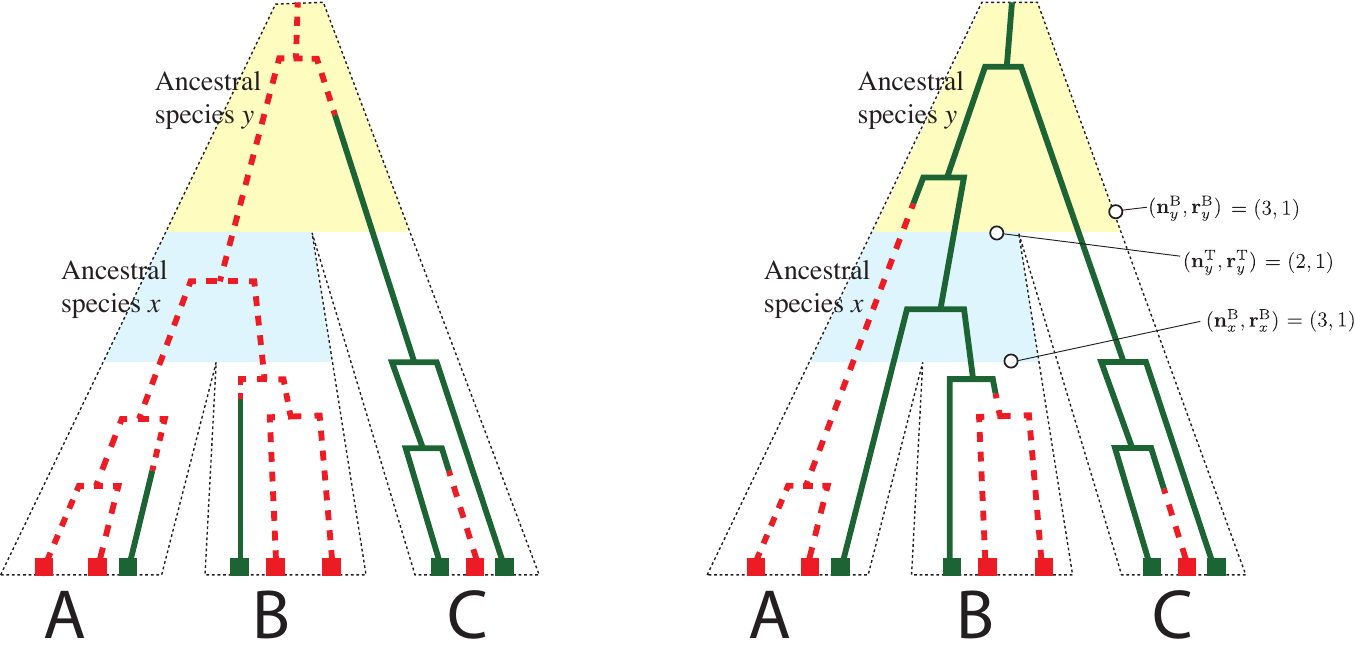} 
   \caption{\small Gene trees in species trees. Each branch in the species trees corresponds to a species that is either contemporary (A,B,C) or ancestral ($x,y$).  The present-day samples are represented by coloured squares along the lower edge of the tree; their colours denote the sampled alleles.  The red (dashed) and green (solid) lines trace out two possible gene trees for these individuals, the red-green colouring indicating which allele is carried by which lineage at any particular time. The random variables $\bNb_x$ and $\bRb_x$ equal the number of lineages, and the number of red lineages, respectively, at the bottom of the branch for ancestral species $x$. The corresponding values at the top of this branch are denoted $\bNt_x$ and $\bRt_x$, respectively.}
   \label{fig:SortingDefinitions}
\end{figure}

Let $x$ be a branch (i.e. ancestral species) in the species tree. Let $\bNb_x$ denote the number of gene tree lineages at the base of the branch $x$. Let $\bNt_x$ denote the number of ancestral lineages at the top of the branch and let $t$ be the length of the branch, measured in units of expected number of mutations (see Figure~\ref{fig:SortingDefinitions}). The minimum possible value for $\bNb_x$ and $\bNt_x$ is one, while the maximum possible value is the total number of individuals sampled in populations at or below $x$, which we denote by $m_x$. The distribution of $\bNt_x$ given $\bNb_x$ is given by the probability in the standard coalescent model of going from $n$ ancestors to $k$ ancestors over time $t$ (measured in units of expected mutations):
\begin{align}
 \Pr[\bNt_x = k | \bNb_x=n] &=  \sum_{r=k}^{n} e^{\frac{-r(r-1)t}{\theta}} \frac{(2r-1)(-1)^{r-k} {k}_{(r-1)} {n}_{[r]}}{k! (r-k)! {n}_{(r)}}, \label{eq:Tavare}  \end{align}
where $n_{[r]} = n(n-1)(n-2)\cdots(n-r+1)$ and $n_{(r)} = n(n+1)\cdots(n+r-1)$,  \citep{Tavare84}. 

When $x$ is an {\em external} branch (adjacent to a leaf) in the species tree, $\bNb_x$ equals the number of samples from the species corresponding to that branch. Let $n_x$ denote this number of samples. Then
\be 
\Pr[\bNb_x \!=\! n] = \begin{cases} 1 & \mbox{if $n\!=\!n_x$;} \\ 0 & \mbox{ otherwise.} \end{cases} \label{eq:countLeaf} 
\ee

Suppose that $x$ is an internal or external  branch in the species tree. Let $m_x$ denote the maximum possible value for $\bNb_x$ or $\bNt_x$, equal to the total number of sampled individuals summed across populations at or below $x$ in the species tree. Suppose that $\Pr[\bNb_x \!=\!k]$ has been computed for all $k$ from $1,2,3,\ldots,m_x$. The value of $\bNt_x$ is determined by the value of $\bNb_x$ using the conditional probabilities in \eqref{eq:Tavare}:
\be  \Pr[\bNt_x \!=\! n] = \sum_{k = n}^{m_x} \Pr[\bNb_x \!=\! k] \Pr[\bNt_x \!=\! n | \bNb_x \!=\! k].  \label{eq:countBranch} \ee

Now suppose that $x$ is an internal branch in the species tree and that  branches $y$ and $z$  are attached to the base of branch $x$. There is no time for coalescent events between the tops of the branches $y$ and $z$ and the bottom of the branch above $x$. 
Hence,  $\bNb_x \!=\! \bNt_y + \bNt_z$ and
\be
\Pr[\bNb_x \!=\! n]  = \sum_{k=0}^n \Pr[\bNt_y\!=\! k] \Pr[\bNt_z\!=\!n-k]. \label{eq:countTrousers}
\ee

Equations \eqref{eq:countLeaf}, \eqref{eq:countBranch} and \eqref{eq:countTrousers} together provide a method for computing $\Pr[\bNb_x \!=\! n]$ for all nodes $x$ in the species tree and all $n \geq 1$. In our implementation, the nodes  are visited in a {\em postorder traversal}, in order from the leaves up to the root, so that a node is always visited {\em after} the required probabilities for the children  have already been computed. When considering a branch attached to a leaf in the species  tree, we use  \eqref{eq:countLeaf} to compute 
$\Pr[\bNb_x \!=\! n]$ for all $n$ and \eqref{eq:countBranch} to compute $\Pr[\bNt_x = n]$ for all $n$. At an internal branch, we use \eqref{eq:countTrousers} to compute $\Pr[\bNb_x = n]$ for all $n$ and \eqref{eq:countBranch} to compute $\Pr[\bNt_x = n]$ for all $n$. 

\subsection*{Computing the partial likelihoods}

We now introduce mutation. We associate the two colours red and green to the two alleles and colour each branch of the gene tree according to the allele state along the branch. Hence a gene tree node will be red or green, depending on whether the corresponding lineage carries a red or green allele at that point. A mutation along a lineage is represented by a change in colour along the branch, and a branch can have  multiple colour changes.

Recall that $\bNb_x$ denotes the number of gene tree lineages at the bottom of a particular branch $x$ in the species tree. We let $\bRb_x$ denote the number of these lineages that carry the red allele at that point, so that $0 \leq \bRb_x \leq \bNb_x$. In the same way, we let $\bRt_x$ denote the number of lineages carrying the red allele at the top of the branch $x$, so that $0 \leq \bRt_x \leq \bNt_x$.

Let $r_z$ denote the number of red alleles in a data set observed in the species associated with an external branch $z$. Our objective is to compute the {\em joint} probability of $(\bRb_z \!=\! r_z)$ over all leaves $z$ in the species tree, conditional on the species tree, sample sizes, and model parameters. To this end, we define a {\em partial likelihood} equal to the conditional likelihood for a subtree of the species tree. Let $\sR_x$ denote the event that 
$(\bRb_z \!=\! r_z)$ holds for every external branch $z$ that is a descendant of branch $x$ in the species tree. That is, $\sR_x$ is short-hand for the event  that the allele counts below $x$ correspond to those observed in the data for a single genetic marker.  For every node $x$ of the species tree, and every choice of $n$ and $r$, we define
 \begin{align}
  \bFb_x(n,r) &=  \Pr[\sR_x | \bNb_x \!=\! n, \bRb_x \!=\! r] \Pr[\bNb_x \!=\!n] \label{eq:bFbDef}
\intertext{and}
   \bFt_x(n,r) &=  \Pr[\sR_x | \bNt_x \!=\! n, \bRt_x \!=\! r]\Pr[\bNt_x \!=\!n] \label{eq:bFtDef}.
   \end{align}
We will see that the values $\bFb_x(n,r)$  and $\bFt_x(n,r)$ can be computed by starting at the leaves and working upwards towards the root, just as in  Felsenstein's likelihood algorithm. Furthermore, when $x$ is the root of the tree, the probability for the entire marker can be determined from the values $\bFb_x(n,r)$. Technically speaking,  $ \bFb_x(n,r)$ is not a partial likelihood; rather it is the product of a partial likelihood ($
 \Pr[\sR_x | \bNb_x \!=\! n, \bRb_x \!=\! r]$) and a probability ($ \Pr[\bNb_x \!=\!n] $). This latter term simplifies the mathematics further on, and makes the computation more numerically stable.  In the same way, $\bFt_x(n,r)$ is the product of a partial likelihood $(\Pr[\sR_x | \bNt_x \!=\! n, \bRt_x \!=\! r])$ and a probability $(\Pr[\bNt_x \!=\!n])$. We will show that these quantities can be computed using dynamic programming and then used to compute the probability of the marker.

 We note that the computation can be extended to multifurcating species trees by converting a multifurcating tree into a bifurcating tree. This is done by replacing any multifurcation with a series of bifurcations, all separated by branches of length $0$. The probabilities of the marker will be unchanged.

\subsubsection*{\em Partial likelihoods for a leaf}

 The simplest case for computing the partial likelihood is when the branch $x$ is attached to a leaf (that is, when $x$ is external). The number of samples from the associated species is $n_x$ and the number of individuals is $r_x$. Hence
 \be
\bFb_x(n,r) = \begin{cases}
    1 & \mbox{ if $n \!=\! n_x$ and $r \!=\! r_x$}\\
    0 & \mbox{ otherwise. } \end{cases} \label{eq:leafL}
\ee

\subsubsection*{\em Partial likelihoods along a branch}

Let $y$ be a branch for which $\bFb_y(n_b,r_b)$ has already been computed, for all $n_b$ and $r_b$. We carefully manipulate the conditional probabilities to obtain an expression for $\bFt_y(n_t,r_t)$. As before, we let $m_y$ denote the number of individuals in sampled from species at or below $y$. Starting with the definition of $\bFt_y$ we have
\begin{align}
\bFt_y(n_t,r_t) &= \Pr[\sR_y | \bNt_y \!=\! n_t, \bRt_y \!=\! r_t ] \Pr[\bNt_y \!=\! n_t] \nn \\
& =  \sum_{n_b \!=\! n_t}^{m_y} \sum_{r_b\!=\!0}^{n_b} \Pr[\sR_y | \bNt_y \!=\! n_t, \bRt_y \!=\! r_t,  \bNb_y \!=\! n_b, \bRb_y \!=\! r_b ] \nn \\
& \quad \quad \times \Pr[\bNb_y \!=\! n_b, \bRb_y \!=\! r_b  | \bNt_y \!=\! n_t, \bRt_y \!=\! r_t ] \Pr[\bNt_y \!=\! n_t].
\end{align}
Recall that $m_y$ denotes the total number of individuals sampled from populations at, or below, branch $y$. We now use the fact that $\sR_y$ is conditionally independent of $\bNt_y$ and $\bRt_y$ given $\bNb_y$ and $\bRb_y$, so that $\Pr[\sR_y | \bNt_y , \bRt_y,  \bNb_y, \bRb_y ] = \Pr[\sR_y | \bNb_y, \bRb_y]$, and
\begin{align}
\bFt_y(n_t,r_t) &=  \sum_{n_b \!=\! n_t}^{m_y} \sum_{r_b\!=\!0}^{n_b} \Pr[\sR_y |\bNb_y \!=\! n_b, \bRb_y \!=\! r_b ] \Pr[\bNb_y \!=\! n_b, \bRb_y \!=\! r_b  | \bNt_y \!=\! n_t, \bRt_y \!=\! r_t ] \Pr[\bNt_y \!=\! n_t] \nn \\
& =  \sum_{n_b = n_t}^{m_y} \sum_{r_b=0}^{n_b} \frac{\bFb_y(n_b,r_b)}{\Pr[\bNb_y \!=\! n_b]} \Pr[\bNb_y \!=\! n_b, \bRb_y \!=\! r_b  | \bNt_y \!=\! n_t, \bRt_y \!=\! r_t ] \Pr[\bNt_y \!=\! n_t]. \nn
\end{align}
After rearranging the conditional probabilities we have 
\[ \Pr[\bNb_y , \bRb_y  | \bNt_y , \bRt_y  ]  = \Pr[ \bRb_y | \bNb_y ,\bNt_y, \bRt_y ] \Pr[  \bNb_y | \bNt_y, \bRt_y ],   \]
which simplifies to $\Pr[ \bRb_y | \bNb_y ,\bNt_y, \bRt_y ] \Pr[  \bNb_y | \bNt_y]$ as the number of red lineages at the top of the branch is conditionally independent of the number of lineages at the bottom, given the number of lineages at the top. Applying Bayes rule
\[ \Pr[  \bNb_y | \bNt_y] \frac{\Pr[\bNt_y]}{\Pr[\bNb_y]} = \Pr[\bNt_y | \bNb_y],\]
we obtain
\begin{align}
\bFt_y(n_t,r_t) &=  \sum_{n_b = n_t}^{m_y} \sum_{r_b=0}^{n_b}  \bFb_y(n_b,r_b)  \Pr[\bRb_y \!=\! r_b| \bNb_y \!=\! n_b, \bNt_y \!=\! n_t, \bRt_y \!=\! r_t ]  \Pr[ \bNt_y \!=\! n_t | \bNb_y \!=\! n_b ] \label{eq:branch}
\end{align}

The term $\Pr[ \bNt_y \!=\! n_t | \bNb_y \!=\! n_b ] $ is evaluated using \eqref{eq:Tavare} above. Computing $\Pr[ \bRb_y | \bNb_y ,\bNt_y, \bRt_y ]$ is more involved. In the special case that $u=v=0$ a closed-form expression exists for this probability \citep{Slatkin96,Nielsen98a,RoyChoudhury08}. To our knowledge, a closed-form expression in the general case has not previously been derived; rather, we express this probability using a matrix exponential. 

Define the matrix $\bQ$ with rows and columns indexed by pairs $(n,r)$ with 
\begin{align}
\bQ_{(n,r);(n,r-1)} & =  (n-r+1)v &\mbox{\hspace{0.5cm}}& 0<r\leq n\nn \\
\bQ_{(n,r);(n,r+1)} & =  (r+1)u && 0 \leq r < n \nn \\
\bQ_{(n,r);(n-1,r)} & =   \frac{(n-1-r)n}{\theta}   && 0 \leq r < n     \label{Qdef}  \\
\bQ_{(n,r);(n-1,r-1)} & =  \frac{(r-1)n}{\theta}   &&0 <r\leq n \nn \\
\bQ_{(n,r);(n,r)} & =  - \frac{n(n-1)}{\theta}  - (n-r)v - ru \nn && 0 \leq r \leq n
\end{align}
and all other entries zero. Here $n$ ranges from $1$ to the number of individuals sampled while for all $n$ we have $0 \leq r \leq n$. Hence $\bQ$ has 
\[\sum_{n=1}^m (n+1) = \frac{1}{2} m(m+3)\]
 rows and columns.
We note that $\bQ$ is not the generator of a process, and the connection with the coalescent and mutation processes is somewhat indirect. The most important feature of the matrix is its role in computing the conditional allele probabilities required for the partial likelihoods:
\begin{theorem} \citep[Theorem 1]{Bryant11cc} \label{thm:transitions}
Suppose that $\bNb$ individuals are sampled from a Wright-Fisher population. Let $\bNt$ denote the number of ancestral lineages at some time $t$ in the past, and let $\bRt$ be the number of these that carry the red allele at that time. Then

\begin{equation}
 \Pr[\bRb \!=\! r | \bNb \!=\! n, \bNt \!=\! n_t, \bRt \!=\! r_t] =  \frac{\exp(\bQ t)_{(n,r);(n_t,r_t) } }{\Pr[\bNt \!=\! n_t | \bNb \!=\! n]}. \label{eq:expQform}
 \end{equation}
\end{theorem}

See \cite{Bryant11cc} for a proof of Theorem~\ref{thm:transitions}. 

\subsubsection*{\em Partial likelihoods at a speciation}

Suppose that a branch $x$ represents a population that diverges into two populations, corresponding to branches $y$ and $z$. Each of the $n$ lineages at the bottom of  branch $x$ came up either from the top of branch $y$ or from the top of branch $z$. If $n_y$ is the number that came up from branch $y$, then $n-n_y$ is the number  from branch $z$. The conditional joint distribution of $\bNt_y$ and $\bNt_z$ is then
\begin{equation}
\Pr[\bNt_y \!=\! n_y, \bNt_z \!=\! n-n_y | \bNb_x \!=\! n] =  \frac{\Pr[\bNt_y \!=\! n_y] \Pr[\bNt_z \!=\! n-n_y]}{\Pr[\bNb_x \!=\! n]}, \label{eq:split1}
\end{equation}
 and is computed using \eqref{eq:countLeaf}, \eqref{eq:countBranch}, and \eqref{eq:countTrousers}. The conditional distribution of red allele counts is given by the hypergeometric distribution:
\begin{equation}
\Pr[\bRt_y \!=\! r_y, \bRt_z \!=\!r\!-\!r_y | \bNt_y \!=\! n_y, \bNt_z \!=\! n-n_y, \bNb_x \!=\! n, \bRb_x \!=\! r]  =   \frac{\binom{n_y}{r_y} \binom{n-n_y}{r-r_y}}{\binom{n}{r}}. \label{eq:split2}
\end{equation}
The value of $n_y$ can range from $n_y = 1$ (one lineage coming from branch $y$) to $n_y = n-1$ (all but one lineage coming from branch $y$.  Combining \eqref{eq:split1} and \eqref{eq:split2}, and summing over $n_y$ and $r_y$, and applying \eqref{eq:bFbDef} and \eqref{eq:bFtDef} we obtain
\beq
\bFb_x(n,r) 
& = &  \sum_{n_y = 1}^{(n-1)} \sum_{r_y = 0}^{r}  \bFt_y(n_y,r_y) \bFt_z(n\!-\!n_y,r\!-\!r_y)  \nn  \frac{\binom{n_y}{r_y} \binom{n-n_y}{r-r_y}}{\binom{n}{r}}       . \nn \\
 \label{eq:trousers}
\eeq
This equation gives the joint probability of the allele counts in the subtree below branch $y$ and the subtree below branch $z$, conditioned on the sum of the respective lineage counts equalling $n$ and the sum of the respective red lineage counts equalling $r$. Note however that the way we have defined $\bFt_y(n,r)$, and $\bFt_z(n,r)$ means these quantities are not partial likelihoods but include also the lineage count probabilities at the nodes (see \eqref{eq:bFbDef} and \eqref{eq:bFtDef}).

\subsection*{Root probabilities}

Let $\rho$ denote the root of the species tree. The probability of the observed data at a genetic marker, conditional on the species tree and model parameters, is
\begin{align}
\Pr[\sR_\rho] & =   \sum_{n = 1}^{m_\rho} \sum_{r = 0}^n \Pr[\sR_\rho | \bNb_\rho = n, \bRb_\rho = r] \Pr[\bNb_\rho = n] \Pr[\bRb_\rho = r | \bNb_\rho = n] \nn \\
&= \sum_{n=1}^{m_\rho} \sum_{r=0}^n \bFb_\rho(n,r)  \Pr[ \bRb_\rho \!=\! r | \bNb_\rho \!=\! n].  \label{eq:like}
\end{align}
Here, $m_\rho$ is the total number of individuals sampled. The term $\Pr[ \bRb_\rho \!=\! r | \bNb_\rho \!=\! n]$ requires some assumptions about what happens in the population above the root. 

Using diffusion models, it can be shown that the allele frequencies in a single population 
have an approximately beta distribution (see, e.g. \citet{Ewens04} pg 174), and this is the distribution used for the root allele probabilities in  \cite{RoyChoudhury08}. It was shown in \cite{Bryant11cc} that exact probabilities under the coalescent model can be derived from Theorem~\ref{thm:transitions} above. 

Let $\bN$ and $\bR$ be the number of lineages and red lineages sampled from a single population of constant size. Let $\bQ$ be the matrix defined in \eqref{Qdef} and let $\bx$ be be the non-zero solution for $\bQ \mathbf{x} = \mathbf{0}$ such that $\mathbf{x}_{(1,0)} + \mathbf{x}_{(1,1)} = 1$. The vector $\bx$ is indexed by pairs in the same way as $\bQ$. Theorem 2 of \cite{Bryant11cc} states then gives $\Pr[\bR \!=\! r | \bN \!=\! n]  =   \mathbf{x}_{(n,r)}$ for all $n$ and and $r$. The matrix $\bQ$ is highly structured and $\mathbf{x}$ can be computed using a simple recurrence in $O(m^2)$ time, where $m$ is the maximum value for $n$.

\begin{figure*}[ht]
\begin{tabular}{|l|}
\hline
{\bf Algorithm} snappLikelihood\\
\parbox{7in}{
\begin{tabbing}
XXX\=XXX\=XXX\=XXX\=XXX\=XX\=XX\= \kill 
{\em Computes the log-likelihood for biallelic data at a genetic marker}\\
\hspace*{6in}\\
{\bf for} each branch $x$ of the species tree in a {\em post-order} traversal\\
\> compute $\Pr[\bNb_x = n]$ for all $n$, using \eqref{eq:countLeaf} if $x$ is external and \eqref{eq:countTrousers} otherwise.\\
\> {\bf if} not at the root, compute $\Pr[\bNt_x = n]$ for all $n$ using \eqref{eq:countBranch}.\\
{\bf end(for)}\\
compute $\Pr[\bR_\rho = r|\bN_\rho = n]$ for all $n,r$ using Theorem 2 of \cite{Bryant11cc}\\
{\bf for} each marker $i$\\
\> {\bf for} each branch $x$ of the species tree in a {\em post-order} traversal\\
\> \> compute $\bFb_x(n,r)$ for all $n,r$, using \eqref{eq:leafL} if $x$ is external and \eqref{eq:trousers} otherwise.\\
\> \> {\bf if} not at the root, compute $\bFt_x(n,r)$ for all $n,r$ using \eqref{eq:branch}.\\
\> {\bf end(for)}\\
\> compute $L_i = \Pr[\sR_\rho] $ using \eqref{eq:like}.\\
{\bf end(for)} \\
{\bf return} $\sum_i \log(L_i)$.
\end{tabbing}}\\
\hline
\end{tabular}
\caption{ \label{fig:algo} High-level outline of the algorithm to compute the log-likelihood of a set of unlinked biallelic markers, given the species tree. A branch $x$ in the species tree is external if it is adjacent to a leaf, otherwise it is internal. In \eqref{eq:countTrousers} and \eqref{eq:trousers} we use $y$ and $z$ to denote the branches attached to the base of branch $x$. }
\end{figure*}

\subsection*{Time complexity}

Fig.~\ref{fig:algo} gives a high-level description of the algorithm for computing the likelihood of a species tree given a collection of unlinked biallelic markers. The time complexity of the algorithm is dominated by two calculations. The first is the evaluation of $\bFb_x(n,r)$ for all $n,r$ using \eqref{eq:trousers}. A direct implementation of the formula would require $O(n^4)$ time per marker, per branch in the species tree, where $n$ is the number of individuals sampled. However the application of two-dimensional convolution algorithms reduces this complexity to $O(n^2 \log n)$ by using the fast Fourier transform; see \cite{Bracewell00}.

The second time-consuming calculation is the computation of $\exp(\bQ t)$, which is  required for the application of \eqref{eq:branch}. We found that standard diagonalisation techniques were both computationally expensive and numerically unstable. Instead, we use the fact that after rearranging \eqref{eq:branch} we only need to be able to evaluate $\exp(\bQ t) \mathbf{v}$ for different vectors $\mathbf{v}$. For this computation we implemented a Carath\'{e}odory-Fej\'{e}r approximation  based on \cite{Schmelzer07} that runs in $O(n^2)$ time per species tree node. To check numerical accuracy, we also implemented the expokit algorithm of \cite{Sidje98}, which is slower than the method of \cite{Schmelzer07} but has more numerical safety checks.

In summary, the time complexity of our likelihood calculation, per marker, is $O(s n^2  \log n)$, where $n$ is the number of individuals and $s$ is the number of species. We implemented a dynamic cache-based system to store partial likelihood values for different subtrees, and multi-threading to take advantage of parallel computation on multiple core machines or graphics processing units.

\subsection*{The SNAPP sampler}

We implemented our likelihood algorithm as the core of a Markov chain Monte Carlo (MCMC) software package SNAPP, which takes biallelic data (e.g. SNPs or AFLP) at multiple loci in a set of species
and returns samples from the joint posterior distribution of
\begin{enumerate}
\item species phylogenies;
\item (relative) species divergence times;
\item (relative) effective population sizes along each branch.
\end{enumerate}
Note that our method does not sample gene trees; it only samples the species tree and its parameters. 

The software is open source, and is available for download from
\begin{quotation}
 {\tt http://www.maths.otago.ac.nz/software/snapp}
\end{quotation}

We have implemented a range of standard priors in the SNAPP package. 
\begin{enumerate}
\item The stationary allele proportions $\pi_0$ and $\pi_1$ were fixed at the observed proportions of red and green alleles in the data. In our experience, the posterior distribution of $\pi_0$ or $\pi_1$ is tightly peaked at the observed value. These observed proportions also determine mutation rates $u$ and $v$, since we measure time in units of expected mutations. 
\item As in \cite{Drummond07}, we assumed a pure birth (Yule) model for the species tree topology and species divergence times, with a hyper-parameter $\lambda$ equal to the birth rate of the species tree. This hyper-parameter was either fixed or allowed to vary with an (improper) uniform hyper-prior.
\item Following \cite{Rannala03}, we used independent gamma prior distributions for the population size parameters $\theta$. We used fixed values for the shape $\alpha$ and inverse scale $\beta$ parameters for the gamma distribution. 
\end{enumerate}
Later, we list the parameters for the prior distributions that we used for simulations and for the analysis of the {\em Ourisia} data. 

The MCMC proposal functions implemented in SNAPP are standard, and are a subset of those available BEAST \citep{Drummond07}, when sampling from molecular clock trees. Briefly, we implemented moves that raise or lower single nodes in the species tree, a move that swaps subtrees in the species tree, moves that alter $\theta$ values for single or multiple populations, and several moves that alter branch lengths and $\theta$ values simultaneously. See \cite{Drummond02} for a detailed discussion of these moves, and \cite{Huelsenbeck01} for a general overview of the use of MCMC methods to sample phylogenies. The  selection of proposal moves is expected to change frequently as the MCMC algorithm is improved, see the (online) SNAPP manual for details.

The execution of the MCMC, and outputs, are controlled via a BEAST-style XML-file, which can be constructed using a graphical user interface. The user has the option of outputting a range of parameters and statistics based on the chain; many more are available by passing the output tree files through TreeStat\footnote{http://tree.bio.ed.ac.uk/software/treestat/}.  Convergence can be assessed for several statistics (e.g. likelihood values, tree length, tree height, summary $\theta$ values) visually using Tracer \citep{Rambaut07} on multiple chains, and using the Gelman-Rubin diagnostic \citep{Gelman92}. Credibility sets are determined by ranking the trees in the sample by decreasing sample frequency, and keeping as many trees as were necessary to obtain a total of 95\% of the sample (after burn in). 


Note that we report only {\em relative} and not {\em absolute} estimates of $\theta$ values. The reason is that once non-polymorphic sites are removed, the allele frequencies of the polymorphic sites appear to contain little or no information about $\theta$. Under the infinite-sites model, the expressions giving  allele count probabilities for for a polymorphic site do not involve $\theta$. Hence, if we condition on sites or markers being polymorphic, the frequency data provides no information about the absolute value of $\theta$. In fact the proportion of segregating sites is a sufficient statistic for $\theta$ in a single population \citep{RoyChoudhury10} under the infinite-sites model.

The situation might be different for the multi-species coalescent, but the following heuristic argument suggests it is not. One interpretation of the infinite-sites model for polymorphic sites is that we are conditioning on there being exactly one mutation in the gene tree. It follows that the allele probabilities for a polymorphic site are invariant to rescaling the length of all branches in the gene tree by a constant. If we rescale the divergence times and $\theta$ values by a constant, the effect on the gene tree distribution will be the same as simply rescaling all the gene trees by the same constant. As this can have no effect on the marker probabilities, the likelihood function is flat. It is not clear the extent to which the same applies for the finite-sites model, however we would expect similar behaviour between the two when the divergence rates are low.

\subsection*{Simulations}

We have tested the likelihood algorithm and the SNAPP software extensively. The  likelihood algorithm and sampler were originally implemented in C++ and then subsequently re-implemented in Java. Core calculations, such as those required to apply Theorem 1, were independently re-implemented and tested in MATLAB. The likelihood values returned by the algorithm for two species and a small number of individuals were identical to those obtained analytically from gene tree probabilities.

We have implemented  a simulator called SimSnapp which generates biallelic polymorphic markers on a species tree according to the multi-species coalescent. Internally,  SimSnapp  generates a gene tree within the species tree and then evolves the marker along that gene tree. If the marker is not polymorphic, then both gene tree and marker are discarded. We note that MCMC-COAL \citep{Rannala03} and Mesquite \citep{Maddison10} could also have been used to generate gene trees, though the fact that most gene trees are discarded made it more efficient to combine gene tree simulation and character simulation within a single program. 

\begin{figure}[htbp] 
   \centering
   \includegraphics[width=5in]{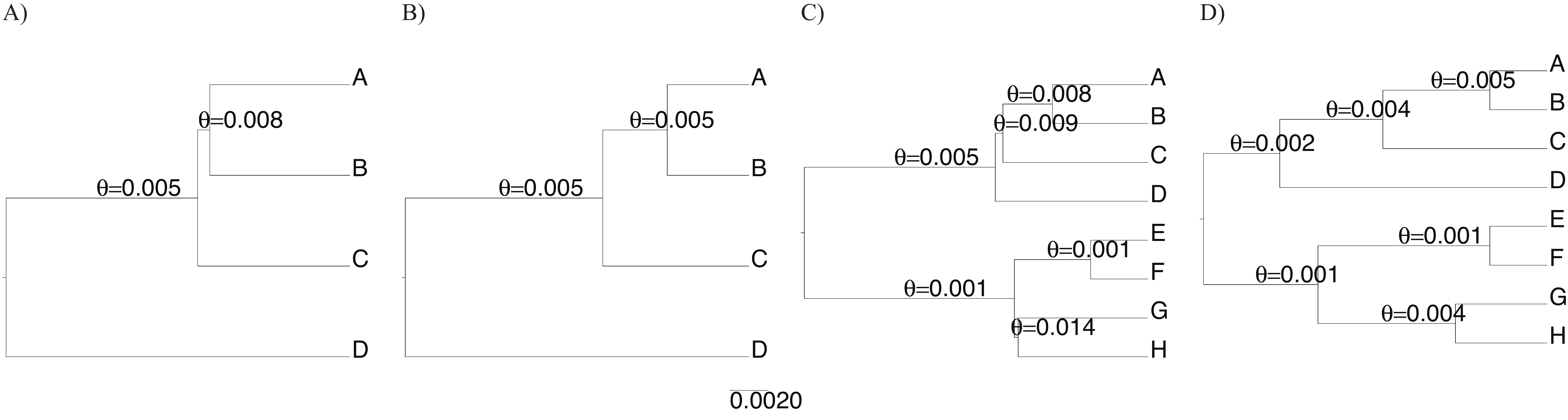} 
   \caption{\small Species trees used for simulations. These trees are identical to those used by \cite{Liu07} to assess the reconstruction of species trees from known gene trees. $\theta$ values are indicated on the tree, and branch lengths are drawn to scale with respect to time. All external branches have $\theta = 0.006$, and branch lengths are drawn to the same scale.  (A) A `hard' four-taxa tree, the difficulty stemming from the short branch separating A, B and C from D. (B) An `easy' four-taxa tree. (C) A `hard' eight-taxa tree, made difficult by the two short branches. (D) An `easy' eight-taxa tree.}
   \label{fig:simTrees}
\end{figure}

In our simulations, we began with the two four-species trees and two eight-species trees used in simulation experiments of \cite{Liu07} Fig.~\ref{fig:simTrees}.  Two of the trees were classified as `hard' due to a short internal branch that would be difficult to resolve: we would expect to need more individuals or more markers to accurately infer these trees. We used the same $\theta$ values as \cite{Liu07} for all internal branches and $\theta=0.006$ for the external branches.

In our first experiment we tested whether, as the number of sites increased, the species tree used to generate data appeared in the credibility set produced by SNAPP. This experiment tests the likelihood algorithm and the sampling algorithm simultaneously. As the number of sites increases, the likelihood function should concentrate around the true value (assuming identifiability) and, consequently, so should the posterior distribution. We also ran SNAPP with four choices of prior. Fo the $\theta$ values we considered a `correct' prior with expectation close to the values used in simulation, and an `incorrect' prior with values averaging 10\% of the true values. In the same way we considered a `correct' prior for species rate corresponding roughly to tree heights of the simulated trees, as well as an `incorrect' prior giving tree heights of around 50\% of the true value.

Chains were started from random trees and initial parameter values drawn from the prior, with chain lengths of 200,000 for the four-taxa case and 400,000 for the eight-taxa case, these lengths being determined by convergence tests on preliminary runs. For this experiment we sampled only one individual per species.

For the second experiment, we examined the posterior distribution of divergence time and $\theta$ values on a fixed tree, using  the `easy' four-taxon tree from before. We simulated ten individuals for each of the four species. We used the same priors as before (`correct' and `incorrect') and generated chains of length one million. For each choice of prior we ran two sampling regimes. For the first scheme, we fixed divergence times (node heights) at their true value and sampled $\theta$ values only. For the second scheme, we sampled divergence times and $\theta$ values at the same time.

\subsection*{Analysis of Ourisia AFLP data}

 \cite{Meudt09} investigated the utility of amplified fragment length polymorphism (AFLP) markers for species delimitation and  reconstruction of evolutionary relationships between New Zealand populations of {\em Ourisia} (Plantaginaceae), the native foxglove. Molecular evidence suggests that {\em Ourisia} species have radiated fairly recently (between 0.4 and 1.3 mya), adapting rapidly to a range of habitats, from sea level to alpine herbfields \citep{Meudt09}. 

 AFLP markers are a readily available source of whole-genome information, well suited to the analysis of closely related species, particularly in the absence of whole-genome sequences (see review in \cite{Meudt07}). 
 The analysis in \cite{Meudt09} used a collection of  2555  non-constant amplified fragment length polymorphism (AFLP) markers for 193 {\em Ourisia} individuals, sampled from 100 locations  in New Zealand and three locations in Australia. Several contrasting tree-based and cluster-based methods were applied, identifying 15 distinct meta-populations. They also detected strong evidence for a split between  a {\em large-leaved group}  and a {\em small-leaved group}, the molecular signal for the split being consistent with differences in both morphology and habitat. The relationship between the species within each of these groups was not well resolved by any method. \cite{Meudt09} argued that this lack of resolution is not due to introgression or insufficient diversity. Two plausible explanations given are the effect of incomplete lineage sorting and the potentially low ratio of phylogenetic signal to noise in  AFLP data.

Here we applied SNAPP to AFLP data from all members of the large-leaved group, producing a data matrix of 69 taxa and 1997 characters. We used a diffuse gamma prior for the $\theta$ values ($\alpha = 10$, $\beta = 100$), with independent $\theta$ values on each branch. We used a pure birth (Yule) prior for the species tree, with birth rate $\lambda$ sampled from an improper uniform hyper-prior. We generated one chain of 1.4 million iterations (sampling every 500 iterations), and two shorter chains (100,000 iterations each) to check convergence. 

\section*{Results}

\subsection*{Simulation experiments: recovering the species tree}

\begin{table}[ht]
\begin{tabular}{r||llll|llll||llll|llll}
\hline
&\multicolumn{8}{c||}{4 taxa} & \multicolumn{8}{c}{8 taxa} \\ \hline
tree	&\multicolumn{4}{c|}{Easy} & \multicolumn{4}{c||}{Hard} &\multicolumn{4}{c|}{Easy} & \multicolumn{4}{c}{Hard} \\
$\theta$-prior & \multicolumn{2}{c}{c}  & \multicolumn{2}{c|}{i} & \multicolumn{2}{c}{c}  & \multicolumn{2}{c||}{i} & \multicolumn{2}{c}{c}  & \multicolumn{2}{c|}{i} & \multicolumn{2}{c}{c}  & \multicolumn{2}{c}{i} \\
$t$-prior 	&c	&i	&c	&i	&c	&i	&c	&i	&c	&i	&c	&i	&c	&i	&c	&i	\\
\hline
100				&1	&1	&1	&1	&3	&3	&3	&3	&1	&1	&1	&1	&22	&21	&12	&14	\\
200				&1	&1	&1	&1	&3	&3	&3	&3	&1	&1	&1	&1	&9	&9	&8	&8	\\
300				&1	&1	&1	&1	&3	&3	&2*	&3	&1	&1	&1	&1	&3*	&3*	&1*	&1*	\\		
400				&1	&1	&1	&1	&3	&3	&3	&3	&1	&1	&1	&1	&9	&9	&8	&8	\\	
500				&1	&1	&1	&1	&3	&3	&3	&3	&1	&1	&1	&1	&6	&6	&4	&5	\\	
600				&1	&1	&1	&1	&3	&3	&3	&3	&1	&1	&1	&1	&8	&8	&6	&6	\\
700				&1	&1	&1	&1	&3	&3	&3	&3	&1	&1	&1	&1	&7	&6	&4	&4	\\
800				&1	&1	&1	&1	&3	&3	&3	&3	&1	&1	&1	&1	&5	&5	&3*	&3*	\\
900				&1	&1	&1	&1	&2	&2	&2	&2	&1	&1	&1	&1	&4	&4	&3	&3	\\
1000			&1	&1	&1	&1	&3	&3	&3	&3	&1	&1	&1	&1	&8	&9	&8	&8	\\
10000			&1	&1	&1	&1	&1	&1	&1	&1	&1	&1	&1	&1	&3	&3	&5	&5	\\
100000			&1	&1	&1	&1	&1	&1	&1	&1	&1	&1	&1	&1	&3	&3	&3	&3	\\
1000000			&1	&1	&1	&1	&1	&1	&1	&1	&1	&1	&1	&1	&2	&2	&2	&2	
\end{tabular}
\caption{\label{table:simulation1} The size of the credibility sets in the first simulation. `tree' indicates which of the trees in Fig.~\ref{fig:simTrees} was used to generate data. `c' and `i' indicate whether `correct' or `incorrect' priors were used on the $\theta$ values and on the speciation rate. Numbers 100 to 1000000 indicate the number of polymorphic sites generated. Values in the table are the numbers of trees in the credibility set. The seven instances where the true tree was not contained within this set are marked by an asterix (*). }
\end{table}

For the first experiment, we  tested whether  SNAPP would recover the tree used to generate the data. We  expected the true tree would be in the 95\% credibility set for almost all replicates, and that the size of the credibility set would shrink as the number of sites increased.

{\em Four taxa with an `easy' tree}: In all simulations, the 95\% credibility set contained the true tree and no other trees. \\
{\em Four taxa with a `hard' tree}: the true tree was  in the 95\% credibility set for all except one instance. The credibility set contained three trees (the three resolutions of the short branch) for data sets with  100 to 1000 sites. The credible set shrunk to one (true) tree with 10,000 or more sites. The two priors had little effect on outcomes.\\
{\em Eight taxa with an `easy' tree}: In all simulations, the 95\% credibility set contained the true tree and no other trees. \\
{\em Eight taxa with a `hard' tree}: The true tree was  in the 95\% credibility set for all except two simulated data sets of 300 and 800 sites.  Otherwise the true tree was contained in the credibility set. There were at least three trees in the credibility set even for data sets with one million sites: SNAPP was unable to resolve the short edge in the species tree. 

Overall, the credibility sets contained the true trees in nearly all the experiments, which is a good indication that the likelihood computation is working correctly.

\subsection*{Simulation experiments- recovering parameters}

\begin{figure}[htb]
 \includegraphics[width=6in]{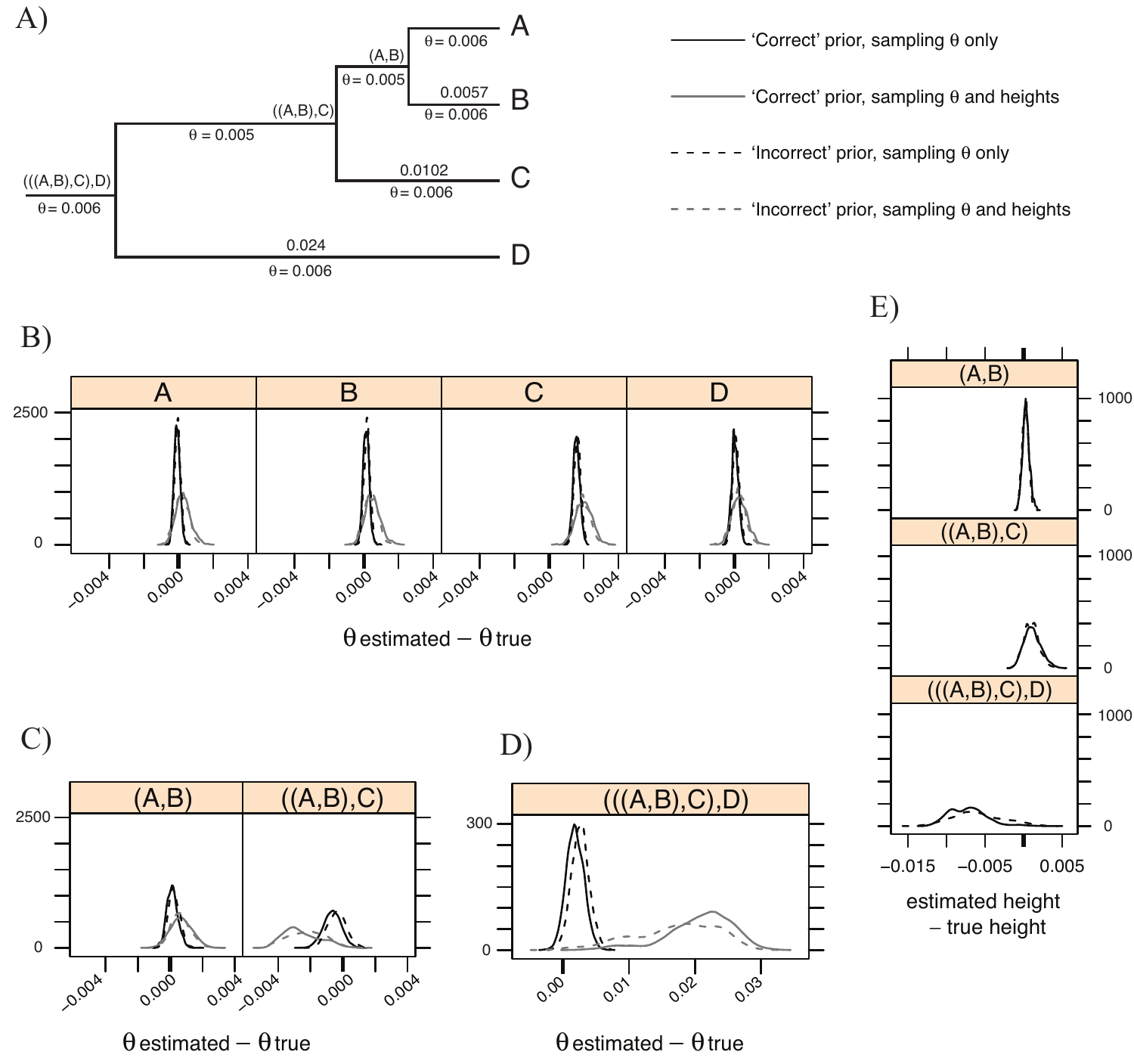} 
\caption{ \label{fig:fourTaxaFig} Posterior distribution for $\theta$ and node heights, simulated data on a four taxon tree.  $\theta$ values are offset by subtracting off the corresponding `true' value on the model tree. Solid lines indicate used of `correct' prior; dashed lines indicate use of `incorrect' prior. Dark lines denote density when only $\theta$ values sampled; gray lines denote density when both divergence times and $\theta$ values sampled. A) The model tree used for the second simulation; B) posterior distribution for $\theta$ on external branches; C) posterior distribution for $\theta$ on internal branches; D) posterior distribution for $\theta$ at root; E) posterior distribution for node heights.}
\end{figure}

The marginal posterior distributions for the node height and $\theta$ parameters appear in Fig.~\ref{fig:fourTaxaFig}.  Once again, the data were analysed using two different sets of priors. The corresponding posterior densities are represented by solid and dashed lines in the figure, and it is apparent that at least in this case, there is very little difference when using the `correct' or `incorrect' prior.

As well as varying the prior, we varied the selection of variables being sampled. This can be used to explore covariances in the posterior.
In one case, we sampled $\theta$ values and node heights, and in the other case we fixed node heights at their true value and sampled only $\theta$ values.  
Fixing the heights greatly reduced the posterior variance of the $\theta$ estimates and shifted them closer to the true value. This could well be due to the (approximate) confounding of $\theta$ and time. Simultaneously scaling the $\theta$ values and branch lengths  induces little change in the likelihood. The fact that the posterior distribution of $\theta$ reduces so markedly when heights are fixed indicates that this invariance under scaling is making a substantial contribution to the variance of the posterior. 

If we compare $\theta$ and node height estimates for different parts of the species tree, we see that both sets of posterior distributions become more diffuse as we head away from the leaves of the species tree and towards the root. This is clearly indicated by the density plots of Fig.~\ref{fig:fourTaxaFig}, which depict the density of the sampled $\theta$ values minus their true values (the {\em offset}).  The $\theta$ distributions for the four observed populations (A,B,C,D in Fig.~\ref{fig:fourTaxaFig}) are comparatively peaked and close to the true value (zero-offset in the figure). In contrast, the $\theta$ distribution for the root is extremely diffuse, especially when we do not fix the node heights. In the same way, the estimation of the height for the root node for these examples appears much more difficult than for the node closest to the leaves. 

If we are to make reliable estimates of $\theta$ values for an ancestral population, we need coalescent events to occur along the corresponding branch. When simulating the data, we recorded the numbers of lineages at each node in the species tree. There were 10 lineages sampled from each population (A), (B), (C), (D). There were, on average, 2.8 lineages at the base of population (A,B), 2.3 lineages at the base of (A,B,C) and 2.0 lineages at the base of the root population. Hence, there were very few coalescent events along internal branches of the species tree, explaining the more diffuse posterior distributions for the corresponding $\theta$ values.

\subsection*{{\em Ourisia} data}

We carried out 1.5 million iterations of the SNAPP MCMC algorithm in approximately 80 hours of computing time  on an Opteron 24 core desktop computer. Additional chains were run to test convergence.

\begin{figure}[htb]
\centerline{
\includegraphics[width=5in]{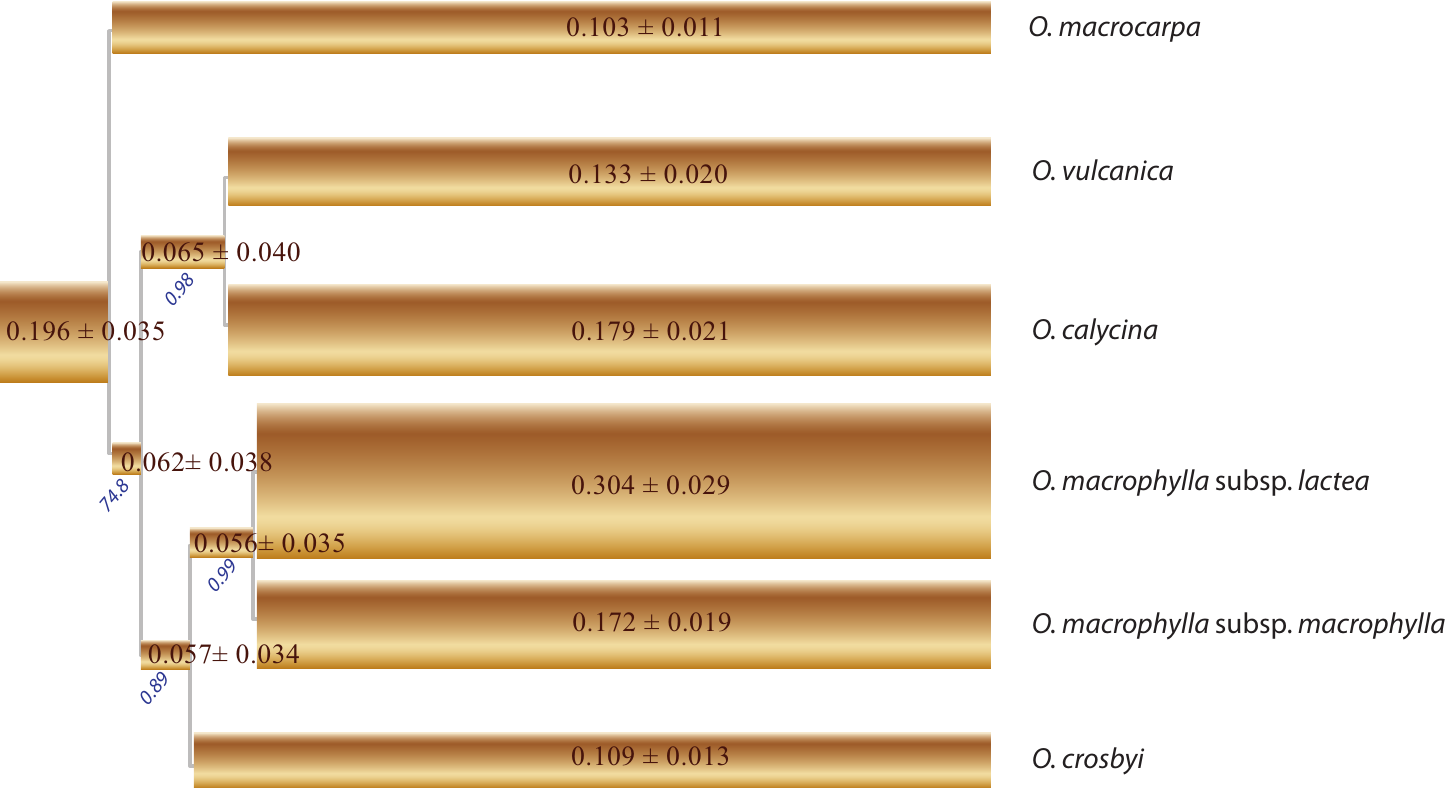} 
}
\caption{ \label{fig:pipetree} Species tree with the highest posterior probability (70\%) for six `Large-leaf' Ourisia species. The thickness of bars proportional to $\theta$ values for the respective populations.  $\theta$ values for each population printed on the pipes. These are relative values only, and were scaled so that sum of $\theta$ values for present day populations is fixed at $1$. The posterior probabilities for internal nodes are printed on an angle.}
\end{figure}

The species tree topology represented in Fig.~\ref{fig:pipetree} has a posterior probability of 70\%. The second most probable topology (15\%) differed only by the position of the root. Interestingly, the AFLP phylogeny of  \cite{Meudt09}, which was obtained using a MrBayes analysis that ignored lineage sorting, had less than 5\% posterior probability. 

SNAPP found significant support for several  relationships between species in this `large-leaved' group. The ({\em O. vulcanica, O. calycina}) clade and the ({\em O. macrophylla, O. crosbyi}) clade both have significant posterior probability. The former clade also appears in the MrBayes tree of \cite{Meudt09}, though the latter does not.

One feature of the tree in Fig.~\ref{fig:pipetree} is that the divergence times for all of the clades are early relative to the age of the tree. This result is consistent with a rapid species radiation at the base of the large-leaved group where an initial swift expansion if followed by a period of consolidation.

The $\theta$ estimates reported in Fig.~\ref{fig:pipetree} are relative estimates only, and many have high posterior variance. One anomaly is the $\theta$ estimate for {\em O. macrophylla} subsp.{\em lactea} which is at least twice that of other species. A Neighbor-Net \citep{Bryant04}  of the AFLP data reveals considerable substructure, and  {\em O. macrophylla} subsp. {\em lactea}  is not monophyletic in neighbor-joining or parsimony analyses \citep{Meudt09}. Hence, the high $\theta$ value could well represent fragmentation within the subspecies or poor delimitation of the subspecies with respect to {\em O. macrophylla} subsp. {\em macrophylla}, rather than a large population. 

In summary, by taking lineage sorting into account, we have been able to extract a well supported phylogenetic tree for {\em Ourisia} species `large-leaved' group. Earlier tree-based and cluster-based analyses were unable to extract such a clear signal from these data \citep{Meudt09}. Our analysis did not support the same tree as a Bayesian phylogenetic analysis which ignored incomplete lineage sorting. Furthermore, our $\theta$ estimates indicate potential fragmentation or poor delimitation in  
 {\em O. macrophylla} subspecies {\em lactea}.

\clearpage

\section*{Discussion}

We have presented a method that takes biallelic markers sampled from multiple individuals from multiple species and computes the likelihood of a species tree topology together with population-genetic  parameters. Our approach implements a full multi-species coalescent model without having to explicitly integrate or sample the gene trees at each loci. With our MCMC sampler, SNAPP, we can concentrate on the parameters of interest: the species tree, population sizes and divergence times, rather than on the problem of traversing through the space of potential gene trees. The likelihood values  we compute are exact (up to numerical error) and do not resort to a simplification or approximation of the full coalescent model. 

Our methods differ from those of \cite{Nielsen98a} and \cite{RoyChoudhury08} by the inclusion of mutation. While mutation is  rare in SNP data from the most closely related populations, it can play a significant role in the evolution of markers for more distant species, for trees with multiple species, or when analysing markers such as AFLP that may have higher mutation rates than SNPs. 

To incorporate  mutation we have derived fundamental new results on the evolution of biallelic markers under the coalescent (Theorems 1 and 2). These formulae extend work of  \cite{Tavare84} on the distribution of ancestral lineage counts, and of \cite{Slatkin96} on the evolution of markers in a population without mutation. It combines the coalescent process, which operates backwards in time, with the mutation process, which works forwards in time. 

The SNAPP sampler complements methods such as  BEST \citep{Liu07} and Star-Beast \citep{Heled10}, which sample  gene trees explicitly. Each is suited to a different kind of data. BEST and Star-Beast can analyse tightly linked markers such as whole gene sequences, whereas SNAPP assumes that all markers are unlinked. It should not be applied to whole genes. Unlike the other methods, SNAPP can analyse tens of thousands of unlinked markers, something that would not be practical if it were necessary to explicitly jointly sample one  gene tree for each marker in a Monte-Carlo algorithm. 

We have reported some of the analyses we have performed to validate the algorithm and our implementation, and to assess the  ability of the method to  infer phylogenetic and demographic parameters. Considerable scope exists for a  more extensive investigation into the strengths, weaknesses and characteristics of the methodology with respect to other approaches. A wide variety of factors affect the  performance of this method, or indeed any method, when inferring trees and parameters. (i) Sufficient mutation must have occurred, but not so much as to cause loss of signal; (ii) $\theta$ values can only be reliably inferred for ancestral populations if there are sufficient coalescent events within these populations; (iii) If $\theta$ values are too high, all coalescences will occur within the root population and no method will be able to infer the tree or $\theta$ values. Alternatively if the values are too low, then all coalescences will occur along pendant branches (as in the examples above), and only patchy information will be available about phylogenetic relationships and population sizes closer to the root of the species tree.  These  issues are likely to be faced not only by SNAPP, but by any method inferring phylogenies and population sizes. 

Because of the potential lack of information about aspects of the species histories, it is of critical importance that methods report not just point estimates, but measures of uncertainty, irrespective of whether they follow Bayesian of frequentist paradigm. Priors either need to be sufficiently diffuse, or  to encode actual information.

We have only briefly discussed the steps taken to improve the computational efficiency of the algorithm and the SNAPP software. A need exists for further improvements in speed,  particularly as the number of individuals in the sample increases. One possibility might be to discard coalescent theory and  revert to diffusion models for allele frequency change \citep{Siren10} as these remove the computational difficulties arising from large numbers of individuals. A hybrid approach is also possible in which computationally efficient continuous models can provide  approximate likelihoods in the SNAPP MCMC algorithm, promising substantial increases in sampling efficiency \citep{Fox08}. An alternative, but intriguing, possibility might be to derive a continuous approximation to the SNAPP likelihood calculation itself, as opposed to approximating the allele frequency changes in the entire population. 

Currently, SNAPP enables a full coalescent analysis of hundreds of thousands of biallelic markers taken from dozens of individuals in multiple species. We see no significant limiting computational factor that, in time, would prevent a SNAPP analysis involving  hundreds of thousands of markers and hundreds, instead of dozens, of individuals.

\section*{Acknowledgments}
 
We thank John Bryant, Alexei Drummond, David Fletcher, Joseph Heled, Pete Lockhart, William Martin, Heidi Meudt, Melanie Pierson and Elisabeth Thompson for help with numerous aspects of this work. DB received funding from the Alexander von Humboldt foundation, the NZ Marsden fund and the Allan Wilson Centre for Molecular Ecology and Evolution. AR was supported in part by NIH program project grant GM-45344, and by NIH Grant R01 GM071639-01A1 (PI: J Felsenstein). JF was funded by National Institutes of
Health grant R01 GM071639, by National Institutes of Health grant
R01 HG004839 (PI: Mary Kuhner), and by interim "life support" funds from the
Department of Genome Sciences, University of Washington. NAR received funding from National Science Foundation grant DBI-1062394.

\pagebreak

\bibliographystyle{abbrvnat}


\begin{thebibliography}{42}
\providecommand{\natexlab}[1]{#1}
\providecommand{\url}[1]{\texttt{#1}}
\expandafter\ifx\csname urlstyle\endcsname\relax
  \providecommand{\doi}[1]{doi: #1}\else
  \providecommand{\doi}{doi: \begingroup \urlstyle{rm}\Url}\fi

\bibitem[Bracewell(2000)]{Bracewell00}
R.~Bracewell.
\newblock \emph{The {F}ourier {T}ransform and its {A}pplications}.
\newblock McGraw Hill, 2000.

\bibitem[Bryant and Moulton(2004)]{Bryant04}
D.~Bryant and V.~Moulton.
\newblock {NeighborNet: An agglomerative algorithm for the construction of
  planar phylogenetic networks}.
\newblock \emph{Mol. Biol. Evol.}, 21:\penalty0 255--265, 2004.

\bibitem[Bryant et~al.(2011)Bryant, RoyChoudhury, Bouckaert, Felsenstein, and
  Rosenberg]{Bryant11cc}
D.~Bryant, A.~RoyChoudhury, R.~R. Bouckaert, J.~Felsenstein, and N.~Rosenberg.
\newblock The conditional coalescent with mutation.
\newblock Submitted to Theoretical Population Biology. ArXiv reference
  submit/0319513, 2011.

\bibitem[Degnan and Rosenberg(2009)]{Degnan09}
J.~H. Degnan and N.~Rosenberg.
\newblock Gene tree discordance, phylogenetic inference and the multispecies
  coalescent.
\newblock \emph{Trends Ecol. Evol.}, 24\penalty0 (6):\penalty0 332--340, 2009.

\bibitem[Degnan and Salter(2005)]{Degnan05}
J.~H. Degnan and L.~A. Salter.
\newblock Gene tree distributions under the coalescent process.
\newblock \emph{Evolution}, 59\penalty0 (1):\penalty0 24--37, 2005.

\bibitem[Drummond and Rambaut(2007)]{Drummond07}
A.~Drummond and A.~Rambaut.
\newblock Beast: {B}ayesian evolutionary analysis by sampling trees.
\newblock \emph{BMC Evol. Biol.}, 7:\penalty0 214, 2007.

\bibitem[Drummond et~al.(2002)Drummond, Nicholls, Rodrigo, and
  Solomon]{Drummond02}
A.~Drummond, G.~Nicholls, A.~Rodrigo, and W.~Solomon.
\newblock Estimating mutation parameters, population history and genealogy
  simultaneously from temporally spaced sequence data.
\newblock \emph{Genetics}, 161\penalty0 (3):\penalty0 1307--1320, 2002.

\bibitem[Efromovich and Kubatko(2008)]{Efromovich08}
S.~Efromovich and L.~S. Kubatko.
\newblock Coalescent time distributions in trees of arbitrary size.
\newblock \emph{Stat. Appl. Genet. Mol. Biol}, 7\penalty0 (1):\penalty0 2, Jan
  2008.

\bibitem[Ewens(2004)]{Ewens04}
W.~J. Ewens.
\newblock \emph{Mathematical Population Genetics}.
\newblock Springer, New York, 2nd edition, 2004.

\bibitem[Felsenstein(1981)]{Felsenstein81}
J.~Felsenstein.
\newblock Evolutionary trees from {DNA} sequences: A maximum likelihood
  approach.
\newblock \emph{J. Mol. Evol.}, 17:\penalty0 368--376, 1981.

\bibitem[Felsenstein(1988)]{Felsenstein88}
J.~Felsenstein.
\newblock Phylogenies and quantitative characters.
\newblock \emph{Annu. Rev. Ecol. Syst.}, 19:\penalty0 445--471, 1988.

\bibitem[Felsenstein(2004)]{Felsenstein04}
J.~Felsenstein.
\newblock \emph{Inferring Phylogenies}.
\newblock Sinauer Associates Inc, 2004.

\bibitem[Fox(2008)]{Fox08}
C.~Fox.
\newblock {Recent advances in inferential solutions to inverse problems}.
\newblock \emph{Inverse Probl. Sci. En.}, 16\penalty0 (6):\penalty0 797--810,
  2008.

\bibitem[Gelman and Rubin(1992)]{Gelman92}
A.~Gelman and D.~Rubin.
\newblock Inference from iterative simulation using multiple sequences.
\newblock \emph{Stat. Sci.}, 7:\penalty0 457--472, 1992.

\bibitem[Hein et~al.(2005)Hein, Schierup, and Wiuf]{Hein05}
J.~Hein, M.~Schierup, and C.~Wiuf.
\newblock \emph{Gene genealogies, variation and evolution: a primer in
  coalescent theory}.
\newblock Oxford University Press, 2005.

\bibitem[Heled and Drummond(2010)]{Heled10}
J.~Heled and A.~Drummond.
\newblock Bayesian inference of species trees from multilocus data.
\newblock \emph{Mol. Biol. Evol.}, 27\penalty0 (3):\penalty0 570--580, 2010.

\bibitem[Hey and Nielsen(2007)]{Hey07}
J.~Hey and R.~Nielsen.
\newblock Integration within the felsenstein equation for improved {Markov
  chain Monte Carlo} methods in population genetics.
\newblock \emph{Proc. Natl. Acad. Sci.}, 104\penalty0 (8):\penalty0 2785--2790,
  2007.

\bibitem[Hudson(1983)]{Hudson83}
R.~R. Hudson.
\newblock Testing the constant-rate neutral allele model with protein sequence
  data.
\newblock \emph{Evolution}, 37:\penalty0 203--217, 1983.

\bibitem[Huelsenbeck et~al.(2001)Huelsenbeck, Ronquist, Nielsen, and
  Bollback]{Huelsenbeck01}
J.~Huelsenbeck, F.~Ronquist, R.~Nielsen, and J.~Bollback.
\newblock Bayesian inference of phylogeny and its impact on evolutionary
  biology.
\newblock \emph{Science}, 294\penalty0 (5550):\penalty0 2310--2314, 2001.

\bibitem[Kubatko et~al.(2009)Kubatko, Carstens, and Knowles]{Kubatko09}
L.~Kubatko, B.~Carstens, and L.~Knowles.
\newblock Stem: species tree estimation using maximum likelihood for gene trees
  under coalescence.
\newblock \emph{Bioinformatics}, 25\penalty0 (7):\penalty0 971--973, 2009.

\bibitem[Liu and Pearl(2007)]{Liu07}
L.~Liu and D.~K. Pearl.
\newblock Species trees from gene trees: Reconstructing {B}ayesian posterior
  distributions of a species phylogeny using estimated gene tree distributions.
\newblock \emph{Syst. Biol.}, 56\penalty0 (3):\penalty0 504--514, 2007.

\bibitem[Maddison and Maddison(2010)]{Maddison10}
W.~P. Maddison and D.~R. Maddison.
\newblock Mesquite: a modular system for evolutionary analysis. version 2.74,
  2010.
\newblock URL \url{http://mesquiteproject.org}.

\bibitem[Meudt and Clarke(2007)]{Meudt07}
H.~Meudt and A.~Clarke.
\newblock {Almost forgotten or latest practice? AFLP applications, analyses and
  advances}.
\newblock \emph{Trends Plant Sci.}, 12\penalty0 (3):\penalty0 106--117, 2007.

\bibitem[Meudt et~al.(2009)Meudt, Lockhart, and Bryant]{Meudt09}
H.~Meudt, P.~Lockhart, and D.~Bryant.
\newblock Species delimitation and phylogeny of a new zealand plant species
  radiation.
\newblock \emph{BMC Evol. Biol.}, 9:\penalty0 111, 2009.

\bibitem[Nei(1987)]{Nei87}
M.~Nei.
\newblock \emph{Molecular Evolutionary Genetics}.
\newblock Columbia University Press, 1987.

\bibitem[Nielsen et~al.(1998)Nielsen, Mountain, Huelsenbeck, and
  Slatkin]{Nielsen98a}
R.~Nielsen, J.~Mountain, J.~Huelsenbeck, and M.~Slatkin.
\newblock {Maximum-likelihood estimation of population divergence times and
  population phylogeny in models without mutation}.
\newblock \emph{Evolution}, 52\penalty0 (3):\penalty0 669--677, 1998.

\bibitem[Pamilo and Nei(1988)]{Pamilo88}
P.~Pamilo and M.~Nei.
\newblock Relationships between gene trees and species trees.
\newblock \emph{Mol. Biol. Evol.}, 5\penalty0 (5):\penalty0 568--83, 1988.

\bibitem[Rambaut and Drummond(2007)]{Rambaut07}
A.~Rambaut and A.~Drummond.
\newblock Tracer v1.4.
\newblock Available from http://beast.bio.ed.ac.uk/Tracer, 2007.

\bibitem[Rannala and Yang(2003)]{Rannala03}
B.~Rannala and Z.~Yang.
\newblock Bayesian estimation of species divergence times and ancestral
  population sizes using {DNA} sequences from multiple loci.
\newblock \emph{Genetics}, 164\penalty0 (4):\penalty0 1645--56, 2003.

\bibitem[RoyChoudhury(2006)]{RoyChoudhury06}
A.~RoyChoudhury.
\newblock \emph{Likelihood Inference for Population Structure, using the
  coalescent}.
\newblock PhD thesis, University of Washington, Aug 2006.

\bibitem[RoyChoudhury and Wakeley(2010)]{RoyChoudhury10}
A.~RoyChoudhury and J.~Wakeley.
\newblock Sufficiency of the number of segregating sites in the limit under
  finite-sites mutation.
\newblock \emph{Theoretical Population Biology}, 78\penalty0 (2):\penalty0
  118--122, 2010.

\bibitem[RoyChoudhury et~al.(2008)RoyChoudhury, Felsenstein, and
  Thompson]{RoyChoudhury08}
A.~RoyChoudhury, J.~Felsenstein, and E.~Thompson.
\newblock A two-stage pruning algorithm for likelihood computation for a
  population tree.
\newblock \emph{Genetics}, 180\penalty0 (2):\penalty0 1095--1105, 2008.

\bibitem[Schmelzer and Trefethen(2007)]{Schmelzer07}
T.~Schmelzer and L.~Trefethen.
\newblock Evaluating matrix functions for exponential integrators via
  carath\'eodory-fej\'er approximation and contour integrals.
\newblock \emph{Electron. T. Numer. Ana.}, 29:\penalty0 1--18, 2007.

\bibitem[Sidje(1998)]{Sidje98}
R.~Sidje.
\newblock Expokit: a software package for computing matrix exponentials.
\newblock \emph{ACM T. Math. Software}, 24\penalty0 (1):\penalty0 130--156,
  1998.

\bibitem[Siren et~al.(2010)Siren, Marttinen, and Corander]{Siren10}
J.~Siren, P.~Marttinen, and J.~Corander.
\newblock {Reconstructing Population Histories from Single Nucleotide
  Polymorphism Data}.
\newblock \emph{Mol. Biol. Evol.}, 28\penalty0 (1):\penalty0 673--683, 2010.

\bibitem[Slatkin(1996)]{Slatkin96}
M.~Slatkin.
\newblock Gene genealogies within mutant allelic classes.
\newblock \emph{Genetics}, 143:\penalty0 579--587, 1996.

\bibitem[Tajima(1983)]{Tajima83}
F.~Tajima.
\newblock Evolutionary relationship of {DNA} sequences in finite populations.
\newblock \emph{Genetics}, 105:\penalty0 437--460, Jan 1983.

\bibitem[Takahata(1989)]{Takahata89}
N.~Takahata.
\newblock Gene genealogy in three related populations: consistency probability
  between gene and population trees.
\newblock \emph{Genetics}, 122\penalty0 (4):\penalty0 957--66, Aug 1989.

\bibitem[Takahata and Nei(1985)]{Takahata85}
N.~Takahata and M.~Nei.
\newblock Gene genealogy and variance of interpopulational nucleotide
  differences.
\newblock \emph{Genetics}, 110\penalty0 (2):\penalty0 325--44, Jun 1985.

\bibitem[Tavar{\'e}(1984)]{Tavare84}
S.~Tavar{\'e}.
\newblock Line-of-descent and genealogical processes, and their applications in
  population genetics models.
\newblock \emph{Theor. Popul. Biol.}, 26\penalty0 (2):\penalty0 119--164, 1984.

\bibitem[Wakeley(2009)]{Wakeley09}
J.~Wakeley.
\newblock \emph{{Coalescent theory: an introduction}}.
\newblock Roberts and Company, Greenwood village, Colorado, 2009.

\bibitem[Wilson et~al.(2003)Wilson, Weale, and Balding]{Wilson03}
I.~Wilson, M.~Weale, and D.~Balding.
\newblock Inferences from {DNA} data: population histories, evolutionary
  processes and forensic match probabilities.
\newblock \emph{J. R. Statist. Soc. A}, 166\penalty0 (2):\penalty0 155--201,
  2003.

\end{thebibliography}

\end{document}